\documentclass[onecolumn]{aastex62}
\usepackage{graphicx,subfigure}
\usepackage{natbib}
\usepackage{comment}
\usepackage{varwidth}
\usepackage{booktabs}
\usepackage{enumitem}
\usepackage{array}

\newcommand{\Msun}{$M_{\odot}$}

\newcommand{\maspy}{$\mathrm{mas~yr^{-1}}$}
\newcommand{\mjypb}{$\mathrm{mJy~beam^{-1}}$}
\newcommand{\kmps}{$\mathrm{km~s^{-1}}$}
\definecolor{mygreen}{rgb}{0.19,0.55,0.11}

\newcommand{\dnsa}{PSR~J0509+3801}
\newcommand{\dnsb}{PSR~J1930$-$1852}
\newcommand{\dnsA}{J0509+3801}
\newcommand{\dnsB}{J1930$-$1852}
\newcommand{\ibcaa}{NVSS~J051132$+$380913} 
\newcommand{\ibcab}{NVSS~J050940$+$381301} 
\newcommand{\ibcac}{NVSS~J050958$+$381358} 
\newcommand{\ibcba}{NVSS~J193138$-$185258} 
\newcommand{\ibcbb}{NVSS~J193102$-$185830} 
\newcommand{\ibcaA}{J051132} 
\newcommand{\ibcaB}{J050940} 
\newcommand{\ibcbA}{J193138} 

\newcommand{\multilinecomment}[1]{}

\begin{document}
\title{VLBA Astrometry of the Galactic Double Neutron Stars \dnsa\ and \dnsb: \\ A Preliminary Transverse Velocity Distribution of Double Neutron Stars and Its Implications}

\author[0000-0002-9174-638X]{Hao~Ding}
\altaffiliation{EACOA Fellow}
\affiliation{Mizusawa VLBI Observatory, National Astronomical Observatory of Japan, 2-12 Hoshigaoka-cho, Mizusawa, Oshu, Iwate 023-0861, Japan}

\author[0000-0001-9434-3837]{Adam~T.~Deller}
\affiliation{Centre for Astrophysics \& Supercomputing, Swinburne University of Technology, PO Box 218, Hawthorn, Victoria 3122, Australia}

\author[0000-0002-1075-3837]{Joseph~K.~Swiggum}
\affiliation{Center for Gravitation, Cosmology, and Astrophysics,
  Department of Physics, University of Wisconsin-Milwaukee, PO Box
  413, Milwaukee, WI, 53201, USA}
\affiliation{Dept. of Physics, 730 High St., Lafayette College, Easton, PA 18042, USA}

\author[0000-0001-5229-7430]{Ryan~S.~Lynch}
\affiliation{Green Bank Observatory, PO Box 2, Green Bank, WV, 24944, USA}

\author[0000-0002-2878-1502]{Shami~Chatterjee}
\affiliation{Cornell Center for Astrophysics and Planetary Science, Ithaca, NY 14853, USA}

\author[0000-0002-3865-7265]{Thomas~M.~Tauris}
\affiliation{Dept. of Materials and Production, Aalborg University, DK-9220~Aalborg {\O}st, Denmark}

\begin{abstract}
 
The mergers of double neutron stars (DNSs) systems are believed to drive the majority of short $\gamma$-ray bursts (SGRBs), while also serving as production sites of heavy r-process elements. 
Despite being key to i) confirming the nature of the extragalactic SGRBs, ii) addressing the poorly-understood r-process enrichment in the ultra-faint dwarf galaxies (UFDGs), and iii) probing the formation process of DNS systems, the space velocity distribution of DNSs is still poorly constrained due to the small number of DNSs with well-determined astrometry. 
In this work, we determine new proper motions and parallaxes of two Galactic DNSs --- \dnsa\ and \dnsb, using the Very Long Baseline Array, 
and estimate the transverse velocities $v_\perp$ of all the 11 isolated Galactic DNSs having proper motion measurements in a consistent manner.
Our correlation analysis reveals that the DNS $v_\perp$ is tentatively correlated with three parameters: spin period, orbital eccentricity, and companion mass. 
With the preliminary $v_\perp$ distribution, we obtain the following findings. 
Firstly, the refined $v_\perp$ distribution is confirmed to agree with the observed displacements of the localized SGRBs from their host galaxy birth sites. 
Secondly, we estimate that around 11\% and 25\% of DNSs remain gravitationally bound to UFDGs with escape velocities of 15\,\kmps\ and 25\,\kmps\, respectively. Hence, the retained DNSs might indeed be responsible for the r-process enrichment confirmed so far in a few UFDGs. Finally, we discuss how a future ensemble of astrometrically determined DNSs may probe the multimodality of the $v_\perp$ distribution.

\end{abstract}


\section{Introduction}
\label{sec:intro}

\subsection{Pulsars in double neutron star systems}
\label{subsec:DNS_pulsars}

Double neutron stars (DNSs) are valuable ``laboratories'' for probing theories of gravity, close binary star evolution, supernovae (SNe), and unveiling the composition of neutron stars (NSs). 
The gravitational-wave (GW) detection of the DNS merger event GW170817 \citep{Abbott17a}, in conjunction with extensive observations across the electromagnetic spectrum  \citep[e.g.][]{Abbott17,Mooley18}, has offered rich insights into the NS interior \citep{Annala18}. A short $\gamma$-ray burst (SGRB) \citep{Goldstein17} coincident with the GW170817 merger event supports the widely believed association between DNS mergers and SGRBs \citep[e.g.][]{Fox05a,Coward12,OConnor22,Fong22}, although most identified SGRBs are beyond the horizon of current GW detectors.
Additionally, light curve monitoring and analysis after the GW170817 event suggest heavy elements were produced during the DNS merger \citep{Drout17,dhs+19}, which reinforces the belief that DNS mergers are a prime source of r-process elements \citep{Eichler89,Korobkin12}. Nonetheless, it remains uncertain whether DNS mergers dominate the production of r-process elements.

Before the violent merger of a DNS is the phase of steady inspiraling. In this phase, Galactic DNSs that host an observable pulsar (hereafter referred to as DNS pulsars) have been extensively studied \citep[e.g.][]{Hulse75,Stairs02,Faulkner05,Kramer06,Jacoby06,Cameron18,Stovall18} with pulsar timing, a technique that measures and models the pulse time-of-arrivals (ToAs) from a pulsar. 
Despite being in shallower gravitational potentials compared to DNS mergers, DNS pulsars have offered some of the most stringent tests on gravitational theories in the strong-field regime with long-term timing observations \citep[e.g.][]{wex14,Fonseca14,Weisberg16,Ferdman20,Kramer21a}.
These tests are made by comparing observed post-Keplerian (PK) parameters, which quantify effects beyond a simple Keplerian model of motion, to the predictions of a specific gravitational theory, e.g. the general theory of relativity (GR).

There are two distinct channels of DNS formation: the predominant ``isolated" channel and the ``dynamical" channel (see \citealp{tv23} and references therein). The former channel gives rise to field DNSs, namely DNS systems born in isolated environments, while both channels can occur in dense regions such as globular clusters \citep{Zevin19}.
At the time of writing, only 19 known DNS pulsars and 2 suspected ones have been discovered from pulsar surveys \citep{Sengar22}, including two found in globular clusters. 

\subsection{Probing the kinematics of field DNSs with precise astrometry}
\label{subsec:proper_motion_and_parallax_of_DNSs}
Precise proper motion and distance measurements for Galactic DNSs are essential for the studies related to DNSs.
The orbital period derivative $\dot{P}_\mathrm{b}$ (or orbital decay) of a DNS system measured by pulsar timing is a PK parameter biased by the Shklovskii effect \citep{Shklovskii70}, an apparent acceleration due to the pulsar's transverse motion. $\dot{P}_\mathrm{b}$ is also affected by the radial acceleration caused by the overall gravity of the Galaxy (see \citealp{Zhu19} and references therein); this acceleration depends on our distance to the pulsar.
Correcting the measured values for these effects requires an accurate measurement of the distance and proper motion. 
As the precision of the observed orbital decay $\dot{P}_\mathrm{b}^\mathrm{obs}$ improves as $t^{-2.5}$ (where $t$ refers to the observing time), the uncertain Shklovskii contribution to $\dot{P}_\mathrm{b}$ can quickly become the limiting factor of the $\dot{P}_\mathrm{b}$ test on gravitational theories \citep[e.g.][]{Deller18,Ding21a}.

Collectively, the measurements of DNS proper motions and distances allow a sample study of space velocities (also known as peculiar velocities or systemic velocities) of DNSs, which, combined with a DNS delay time distribution and galactic gravitational potentials, can be used to predict the spatial distribution of DNS merger sites within, or nearby, a host galaxy \citep[e.g.][]{bsp99,vt03,fb13,Beniamini16a,tkf+17,Vigna-Gomez18,Andrews19a,Beniamini19a,Zevin22}.
Such a prediction, once made, can be compared to DNS mergers \citep{Abbott17} and SGRBs localized with respect to their host galaxies, in order to probe the poorly defined selection biases (on both sides of the comparison), and identify exotic transient events. 
Most pinpointed SGRBs are found far from the centers of their host galaxies and star-forming regions \citep[e.g.][]{Fox05a,Fong13,Tunnicliffe14,OConnor22,Fong22}. 
Assuming SGRBs are predominantly generated by DNS mergers, the transverse space velocities $v_\perp^\mathrm{SGRB}$ of $\approx20$--140\,\kmps\ with a median of $\approx60$\,\kmps\ were inferred from the SGRB displacements from their expected birth sites \citep{Fong13}, in conjunction with the indicative DNS delay time estimated by \citet{Leibler10}.
On the other hand, as the direct link between SGRB localizations and DNS formation theories, the space velocities of Galactic DNSs have been previously investigated with $\leq9$ proper motion and distance measurements \citep{Wong10,tkf+17,Haniewicz21}: the preliminary observational constraints suggest overall consistency with $v_\perp^\mathrm{SGRB}$.

Furthermore, a space velocity distribution of DNSs is crucial for understanding the source of excess r-process elements in ultra-faint dwarf galaxies (UFDGs; \citealp{Ji16,Hansen17,Hansen20}) where the escape velocities ($\sim15$\,\kmps) can be easily surpassed due to the NS natal kicks received at supernovae \citep{Beniamini16,Safarzadeh19}. On the other hand, if a considerable fraction (e.g., $\gtrsim10$\%) of DNSs have a space velocity of $\leq15$\,\kmps, then it is plausible that the r-process elements detected in UFDGs came from DNS mergers bound by the gravity of the host UFDGs. Otherwise, the r-process elements in UFDGs must be produced through fast-merging DNSs \citep{Safarzadeh19} or other mechanisms.

Finally, the DNS formation is not yet fully understood: for example, it remains debated under which conditions an electron-capture supernova (ECSN), instead of a conventional iron core-collapse supernova (CCSN), can give rise to the second-born NS \citep[e.g.][]{plp+04,pdl+05,kp11,Jones16,Giacobbo19,tj19}. As ECSNe are expected to give smaller natal kicks than CCSNe \citep[e.g.][]{kjh06,dbo+06,Gessner18}, DNS space velocities can be used to probe the formation channels of field DNSs. Specifically, if both the CCSN and ECSN channels contribute to the field DNS productions, one would likely find a bimodal feature in the DNS space velocity distribution. However, the picture is more blurred because i) the narrow progenitor mass range leading to ECSNe depends on metallicity \citep{plp+04,sl18}, and UFDGs are very metal-poor \citep{fws+23}, and ii) ultra-stripped SNe \citep{tlm+13,tlp15}, occurring when the last-formed NS is produced in DNS progenitor systems, may also produce NSs with small kicks in both CCSNe and ECSNe \citep[e.g.][]{tkf+17,mth+19}.
Additionally, a positive correlation has been proposed between the second-born NS masses and SN kick magnitudes, and thus space velocities of DNSs \citep{tkf+17,bwvc24}, and possibly also between second-born NS masses and orbital eccentricities, which can also be examined with more proper motion and distance measurements of field DNSs.

Pulsar timing and VLBI astrometry can both provide high-precision proper motions and potentially distances for DNSs. 
However, the lower timing precision attainable for the moderately recycled pulsars typically found in DNS systems (compared to fully recycled ``millisecond" pulsars with periods of $\lesssim10$\,ms) means that long campaigns -- often 10+ years -- are required to detect proper motion and annual geometric parallax with adequate precision.
VLBI astrometry, on the other hand, generally only requires two well-separated observations to detect a high-precision proper motion, and O(10) observations over a 1 year (or longer) span to measure annual geometric parallax with a precision of tens of microarcseconds, irrespective of the pulse period.

To date, only 5 DNSs have been well measured astrometrically with VLBI \citep{Deller18,Kirsten14,Kramer21a,Ding21a,Ding23}, including 4 field DNSs (i.e. PSR~B1913$+$16, PSR~B1534$+$12, PSR~J0737-3039A/B and PSR~J1518$+$4904) and a DNS in the globular cluster M15.
In this work, we carried out VLBI astrometry on two field DNSs --- \dnsa\ and \dnsb\ (see Section~\ref{sec:inidividual_systems} for their introductions). 
In Sections~\ref{sec:obs} through \ref{sec:astrometric_inference}, we describe the VLBI observation strategy and data reduction procedures. In particular, a novel astrometric tactic utilising multiple reference sources is introduced in Section~\ref{subsec:2Xtactic}. 
We present the calculation of distances and transverse space velocities for 11 field DNSs in Section~\ref{sec:D_and_Vt}. The obtained sample of transverse space velocities is discussed thoroughly in Section~\ref{sec:discussion}, while individual DNS systems are discussed in Section~\ref{sec:inidividual_systems}.
Throughout this paper, uncertainties are provided at 68\% confidence, unless otherwise stated.

\section{Observations}
\label{sec:obs}

\begin{figure*}
  \plottwo{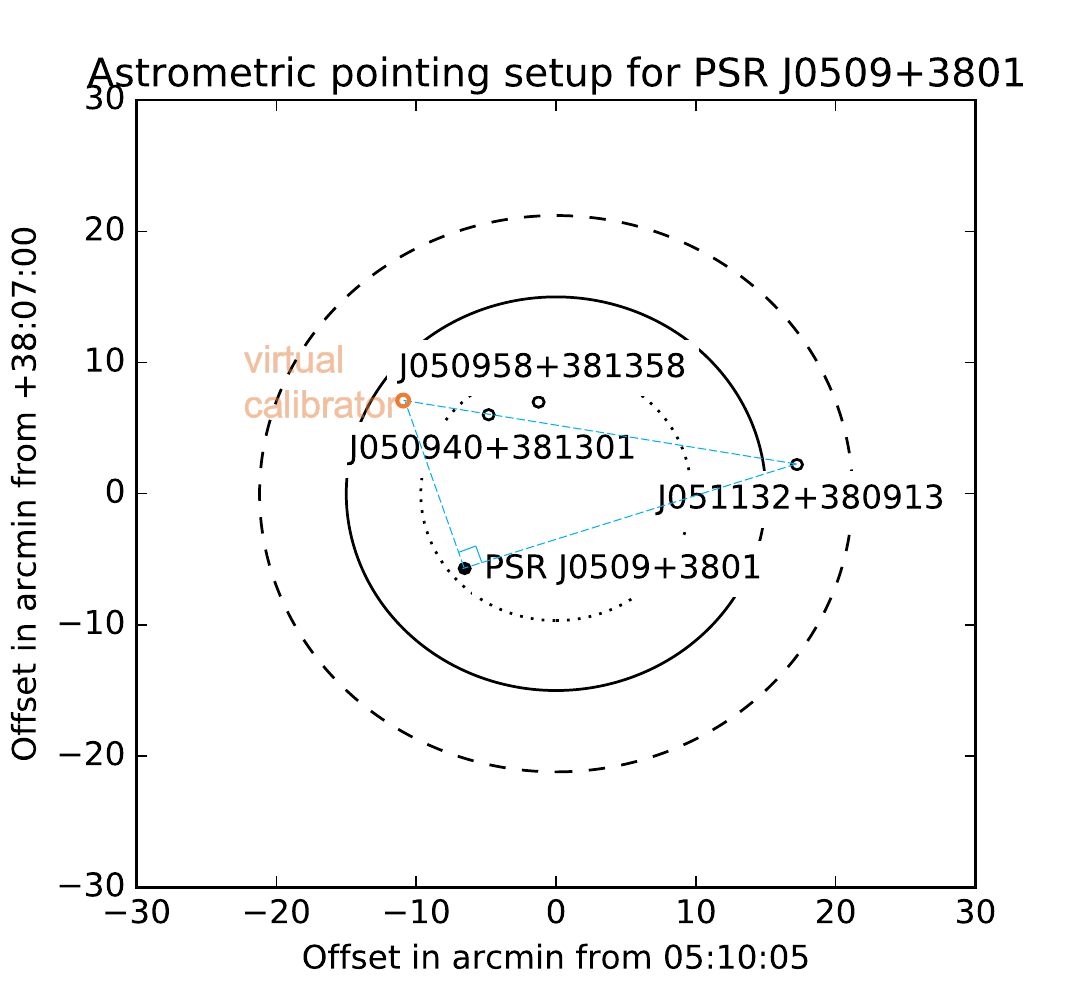}{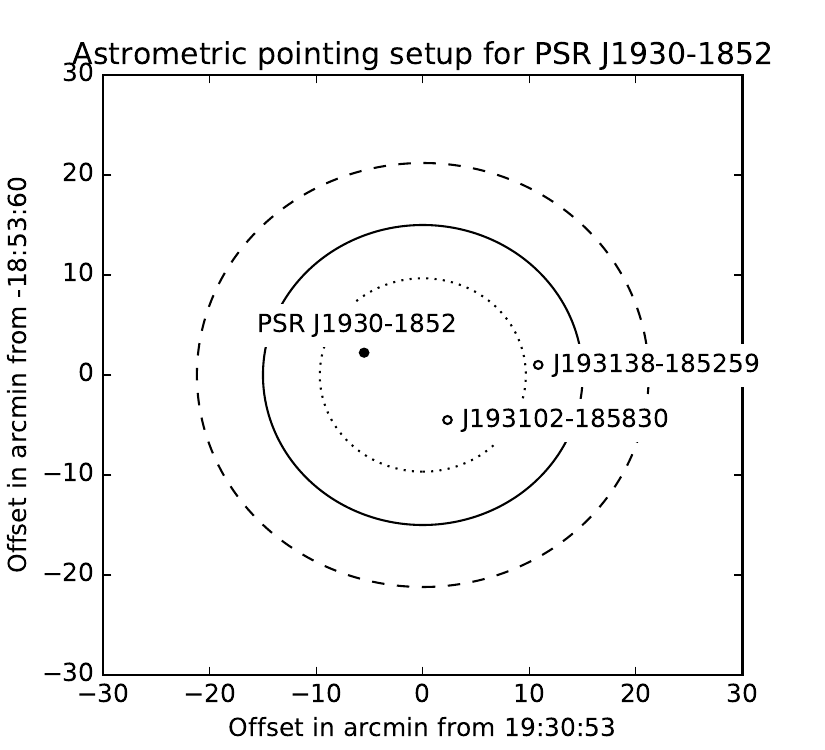}
  \caption{The location of the target pulsar and the background calibrator sources within the VLBA primary beam.  The 75\%, 50\%, and 25\% beam response level is plotted with dotted, solid, and dashed lines respectively. 
  \ibcba\ is used as the phase reference source for \dnsb.
  The astrometry of \dnsa\ is carried out using the novel 2X (read as two cross) strategy (see Section~\ref{subsec:2Xtactic} for more explanations), where the virtual calibrator illustrated as the orange circle is used alongside the primary in-beam calibrator \ibcaa. The virtual calibrator should lie on the line that connects the primary in-beam calibrator \ibcaa\ and the secondary in-beam calibrator \ibcab, and form a 90\degr\ angle with \ibcaa\ at the position of \dnsa. Accordingly, the virtual calibrator is a factor of 1.27 further away from \ibcaa\ compared to \ibcab.}
  \label{fig:pointings}
\end{figure*}

For both \dnsa\ and \dnsb, we made a total of eight observations with the Very Long Baseline Array (VLBA) in the period 2015 March -- 2016 November under the project code BD186.  In each observation, we nodded between the target field (containing the pulsar and one or more extragalactic sources used as ``in-beam" calibrators) and a nearby gain and delay calibrator source.  The location of the pulsars and the background calibrator sources are shown in Figure~\ref{fig:pointings}.  We also included two scans on a bright fringe-finder source, used to calibrate the instrumental bandpass.  Each observation lasted 4 hours, with approximately 2.7 hours spent on the target field.  The target sources and their corresponding calibrators are listed in Table~\ref{tab:sources}.

\begin{deluxetable*}{llcc}
\tabletypesize{\small}
\tablewidth{0.98\textwidth}
\tablecaption{\label{tab:sources}Observed sources}
\tablehead{
\colhead{Source type} & \colhead{Source name} & \colhead{S$_{1.4}$\tablenotemark{A}} &  \colhead{Separation\tablenotemark{B}}  \\
\colhead{}     & \colhead{}   & \colhead{(\mjypb)}                         & \colhead{(arcmin)}     
}
\startdata
Target  			& \dnsa  					& 0.16     & 0 \\ 
Off-beam calibrator  	& ICRF J050905.8+352817	& 174.4     & 153.1 \\
In-beam calibrator 	&  \ibcaa						& 12.1     & 25.1 \\ 
In-beam calibrator  	&  \ibcab  						& 4.9   & 11.9 \\
In-beam calibrator  	&  \ibcac  						& 2.5    & 13.8 \\ 
\\
Target   			& \dnsb 					& 0.12    & 0 \\ 
Off-beam calibrator 	& ICRF J192809.1$-$203543	& 99.0    & 109.1 \\
In-beam calibrator  	&  \ibcba  						&   29.9   & 16.4 \\  
In-beam calibrator  	&  \ibcbb  						&   11.0   & 10.3
\enddata
\tablenotetext{A}{Unresolved flux intensity (median period-averaged flux density in the case of the pulsars) at 1.4 GHz.}
\tablenotetext{B}{Angular separation from the target pulsar.}
\end{deluxetable*}

The instrumental setup used a standard continuum setup for the 20~cm band, with eight subbands of width 32 MHz recorded in dual circular polarisation at 2-bit precision for a total data rate of 2 Gbps per antenna.  Within the frequency tuning limitations afforded by the VLBA's wideband recording system, we attempted to avoid regions of frequency space known to contain strong radio frequency interference at VLBA stations, and selected subbands in the range 1392 -- 1748 MHz.  The first two observations of \dnsb\ yielded only weak detections, leading us to conclude both that the catalog flux density for this pulsar was over-estimated, and that its spectral index was steeper than expected.  Accordingly, for the third observation of \dnsb\ onwards, we changed observing frequencies to span the range 1252 -- 1508 MHz. We note that this change is likely to cause a small (if not negligible) reference point offset for the first two observations of \dnsb\ due to the frequency-dependent core shift \citep[e.g.][]{Voitsik18} of the in-beam calibrator NVSS~J193138$-$185258, and potentially lead to slightly larger scaling factor $\eta_\mathrm{EFAC}$ of the fiducial systematic errors in the astrometry inference (see Section~\ref{sec:astrometric_inference}). Data was correlated using the VLBA DiFX correlator \citep{DiFX2}, making use of pulsar gating to improve the signal--to--noise ratio on the target pulsars. For the pulsar gating, we adopted the pulse ephemerides acquired from pulsar timing observations using the Green Bank Telescope (GBT). 

\section{Data Reduction}
\label{sec:data_reduction}

We make use of the ParselTongue/AIPS \citep{Kettenis06,Greisen03} astrometric reduction pipeline {\tt psrvlbireduce}\footnote{\url{https://github.com/dingswin/psrvlbireduce}} described in \citet{Deller16,Ding23}, and here we briefly summarise the main steps.  {\em A priori} amplitude calibration including correction for the primary beam response is followed by time-independent delay and bandpass calibration using the fringe finder source, and then by time-dependent delay and gain calibration using the out-of-beam calibrator.  The calibrator source models used in these steps, and also in the phase calibration refinement that follows, were derived from a concatenated dataset that including all eight observations, and have been made available online\footnote{\label{footnote:publication_materials}available on Zenodo under an open-source 
Creative Commons Attribution license:
\dataset[doi:10.5281/zenodo.11114889]{https://zenodo.org/doi/10.5281/zenodo.11114889}} to facilitate reproduction of our data reduction results. The cumulative calibration derived so far is then applied to the sources in the target field (the target pulsar and the in-beam calibrators), and the calibrated datasets are split and averaged in frequency.

\subsection{The ``2X'' tactic: doubling the position measurements with two near-field calibrators in perpendicular directions}
\label{subsec:2Xtactic}

The in-beam calibrator data have been used to refine the phase calibration for the target source, eliminating temporal interpolation along with reducing spatial interpolation of calibrator solutions.  
In a normal procedure, self-calibration is performed on one (or sometimes multiple) in-beam calibrator sources  \citep[e.g.][]{Deller19}; the acquired solutions are then applied to all sources in the target field. The multi-source self-calibration technique \citep{Middelberg13,Radcliffe16} is especially useful when all in-beam calibrators are slightly too faint for self-calibration. 
On the other hand, in the opposite scenario where at least two in-beam calibrators are bright enough for self-calibration (or, alternatively, at least two relatively faint in-beam calibrators are identified around a sufficiently bright target; see, e.g., Section~4.1.2 of \citealp{Ding23}), it is possible to further enhance the astrometric precision by doubling the number of relative position measurements. 
Specifically, self-calibration solutions can be derived independently on each sufficiently bright in-beam calibrator. As the self-calibration solutions are applied to the target, the position of the target is measured with respect to the in-beam calibrator. Namely, one target position can be measured with respect to each independently calibrated in-beam calibrator.

However, systematic errors in the position offsets measured against different in-beam calibrators are generally at least partially correlated, due to the predominance of direction-dependent calibration errors in the systematic error budget. As an extreme and straightforward example, two in-beam calibrators situated at the same sky position would render almost perfectly correlated self-calibration solutions.  
Assuming direction-dependent terms dominate the self-calibration solutions, and that these terms change linearly with the sky position of the in-beam calibrator, the solutions obtained with two in-beam calibrators are largely independent of each other only when the two in-beam calibrators are in perpendicular directions as viewed from the target (and even in this case, other sources of systematic error, such as variations taking place between adjacent calibration solutions, remain correlated). Such a pair of in-beam calibrators are hereafter referred to as orthogonal in-beam calibrators, or simply orthogonal calibrators.
The chance of having a pair of physical orthogonal in-beam calibrators is rather small. However, with the help of the 1D interpolation technique \citep[e.g.][]{Fomalont03,Doi06,Ding20c}, the target can be measured with respect to a virtual calibrator (see \citealp{Ding20c} for more explanations), whose effective position can be manipulated (either along the line connecting two self-calibratable in-beam calibrators or along the line connecting the out-of-beam main phase calibrator and a self-calibratable in-beam calibrator) to form an orthogonal in-beam calibrator pair with a physical in-beam calibrator. For brevity, we hereafter refer to this novel strategy of VLBI astrometry as the ``2X'' (read as ``two cross'') strategy.

Interestingly, both DNSs possess at least two self-calibratable in-beam calibrators (see Table~\ref{tab:sources}). Therefore, the 2X strategy is potentially applicable to the astrometry of both DNSs. 
Based on the brightness and compactness of the in-beam calibrators as well as their angular distances from the target, we adopted \ibcaa\ and \ibcba\ as the primary in-beam calibrators for \dnsa\ and \dnsb, respectively. In-beam calibrators other than the primary in-beam calibrators are referred to as secondary in-beam calibrators. 
Unfortunately, both \ibcac\ and \ibcbb\ display a resolved jet feature in the right ascension direction. Any time dependence to the jet brightness profile along  the jet direction may corrupt the parallax measurement \citep[e.g.][]{Deller13}, given that the parallax signature of either DNS is predominantly revealed in the right ascension direction. In such a case, the systematic error budget captured by the self-calibration solution might no longer be dominated by direction-dependent effects, breaking the underlying assumption of the 2X calibration strategy.
Therefore, \ibcac\ and \ibcbb\ are not used as  secondary in-beam calibrators. Accordingly, we only implemented the 2X tactic on \dnsa\ with its primary in-beam calibrator \ibcaa\ and the secondary in-beam calibrator \ibcab.

For each DNS, self-calibration was first performed on the primary in-beam calibrator; the acquired solutions were applied to all sources in the target field. In this way, the target position is measured with respect to the primary in-beam calibrator.
After the application of in-beam phase calibration solutions to all target-field sources, we imaged both the target sources and the in-beam calibrator sources after dividing by the calibrator source model using natural weighting.  
From each image per source per epoch, we extracted a position and uncertainty using the image-plane fitting task {\tt JMFIT} in {\tt AIPS}. The positions so obtained are essentially anchored to the primary in-beam calibrator.

For \dnsa\ only, we went one step further in order to fulfill the 2X strategy: this time \dnsa\ needs to be phase-referenced to a virtual calibrator that forms orthogonal calibrators with the primary in-beam calibrator \ibcaa.
In practice, we implemented 1D interpolation along the line connecting the \ibcaa\ and \ibcab\ (see \citealp{Ding20c} for the detailed procedure), then applied the solutions to only \dnsa. By placing the virtual calibrator 1.27 times further away from \ibcaa\ compared to \ibcab, the virtual calibrator and \ibcaa\ form an orthogonal pair around \dnsa\ (see Figure~\ref{fig:pointings}). In this instance, the virtual calibrator is further away from \dnsa\ than \ibcab\ (which may not be the case given a calibrator plan of another target); this means that the systematic errors (of the target positions) resulting from direction-dependent propagation effects are accordingly amplified. 
Despite this small setback, the overall astrometric precision is expected to be improved (see Section~\ref{sec:astrometric_inference}).

\section{The inference of astrometric parameters}
\label{sec:astrometric_inference}

After the VLBI data reduction (see Section~\ref{sec:data_reduction}), we compiled the positions of \dnsa\ and \dnsb\ across all epochs, which are made available\textsuperscript{\ref{footnote:publication_materials}} as the ``pmpar.in.preliminary'' files. For \dnsb, one series of positions was measured with respect to the primary in-beam calibrator \ibcba, while two position series were acquired for \dnsb\ against two different in-beam calibrators, as the 2X strategy had been applied (see Section~\ref{subsec:2Xtactic}).
The uncertainties of these target positions only reflect the thermal noise in the target images. On top of the random positional errors, we estimated systematic uncertainties using an  empirical estimator (Equation~1 of \citealp{Deller19}) based on the angular separation of the calibrator and target (taking the fiducial value as explained in Section~3 of \citealp{Ding23}), and added them in quadrature to the random errors. The full positional uncertainties are provided online\textsuperscript{\ref{footnote:publication_materials}} in the ``pmpar.in'' files.

\begin{figure}
  \centering
  \begin{minipage}{\textwidth}
    \centering
    \includegraphics[width=0.7\textwidth]{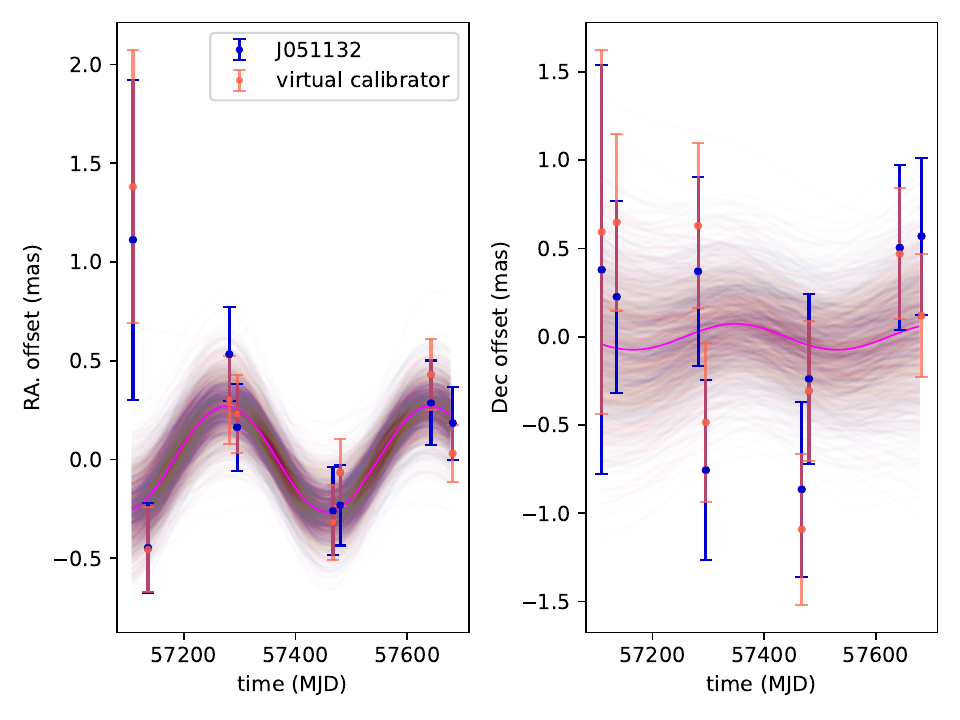}
    
    \textbf{(a)} \dnsa
    \label{fig:J0509_parallax}
  \end{minipage}
  
  \vspace{10pt} 
  
  \begin{minipage}{\textwidth}
    \centering
    \includegraphics[width=0.7\textwidth]{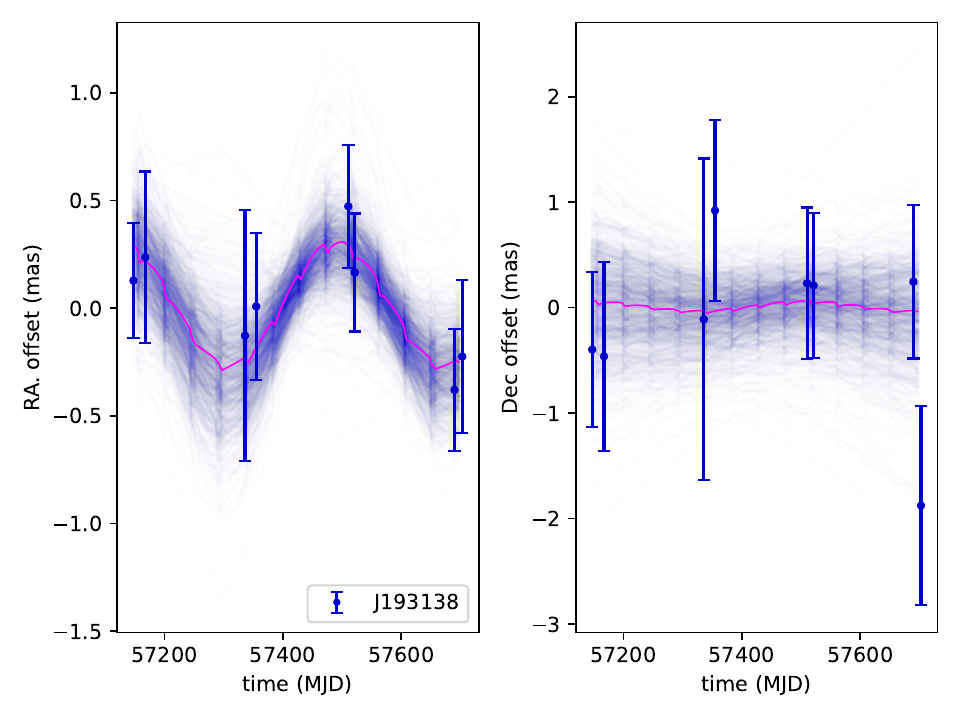}
    
    \textbf{(b)} \dnsb
    \label{fig:J1930_parallax}
  \end{minipage}
  
  \caption{Sky position evolution with time after the removal of proper motion; parallax dominates, while orbital reflex motion is (almost) negligible. The best-fit astrometric model is outlined in pink, while Bayesian simulations are overlaid. 
  The reference sources (see Table~\ref{tab:sources} for their full names) of the position measurements are provided in the legends.
  In particular, the astrometric model of \dnsa\ is inferred from two series of positions measured with respect to two calibrators (see Section~\ref{subsec:2Xtactic}). Orbital parameters (i.e., orbital inclination and ascending node longitude) are only inferred (on top of other astrometric parameters) for \dnsb, causing the small wobbles of the parallax signature in the lower panel (see Section~\ref{sec:astrometric_inference} for more details). }
  \label{fig:parallax_signature}
\end{figure}

Following \citet{Ding23}, we inferred the astrometric parameters of the two DNSs using the astrometric Bayesian inference package {\tt sterne.py}\footnote{\url{https://pypi.org/project/sterne/}} \citep{Ding21a}. The revealed parallax signatures are illustrated in Figure~\ref{fig:parallax_signature}. The inferred values are presented in Table~\ref{tab:astrometric_results}, while the corner plots of the inferences are available online\textsuperscript{\ref{footnote:publication_materials}}. Among the parameters of inference, $\eta_\mathrm{EFAC}$ is the scaling factor on the fiducial systematic error (see Equation~1 of \citealp{Ding23}); $i_\mathrm{orb}$ and $\Omega_\mathrm{orb}$ stand for orbital inclination angle and the ascending node longitude, respectively. As the detectability of orbital motion with VLBI, quantified by $\eta_\mathrm{orb}$ defined with Equation~3 of \citealp{Ding23}, is very low ($\eta_\mathrm{orb}=0.03$, as compared to 1) for \dnsa, we only inferred $i_\mathrm{orb}$ and $\Omega_\mathrm{orb}$ (on top of other astrometric parameters) for \dnsb\ (with $\eta_\mathrm{orb}=1.1$). Nevertheless, without useful prior knowledge of either orbital parameter, we did not achieve non-trivial constraints on $i_\mathrm{orb}$ and $\Omega_\mathrm{orb}$ of \dnsb\ (see Table~\ref{tab:astrometric_results} and the online corner plot\textsuperscript{\ref{footnote:publication_materials}}).

\subsection{The absolute reference positions of \dnsa\ and \dnsb}
\label{subsec:absolute_positions}

We estimated the absolute reference positions of \dnsa\ and \dnsb\ in the same way as described in Section~3.2 of \citet{Ding20}, except that, instead of the bootstrap uncertainties of the relative reference positions with respect to the primary in-beam calibrator, the Bayesian ones were adopted here as part of the error budget of the absolute reference positions. 
At the reference epoch MJD~57381, the absolute position of \dnsa\ is
$\alpha_\mathrm{J0509+3801}=05^{\rm h}09^{\rm m}31\fs 78871\pm0.4\,\mathrm{mas}\pm0.8\,\mathrm{mas}$, $\delta_\mathrm{J0509+3801}=+38\degr01'18\farcs0730\pm0.6\,\mathrm{mas}\pm0.8\,\mathrm{mas}$, while the absolute position of \dnsb\ is $\alpha_\mathrm{J1930-1852}=19^{\rm h}30^{\rm m}29\fs7156\pm2.2\,\mathrm{mas}\pm0.8\,\mathrm{mas}$, $\delta_\mathrm{J1930-1852}=-18\degr51'46\farcs164\pm8.1\,\mathrm{mas}\pm0.8\,\mathrm{mas}$ at MJD~57433. For both positions, the second error terms are indicative of the frequency-dependent core shift of the off-beam phase calibrator (\citealp{Sokolovsky11}, also see the explanation in Section~3.2 of \citealp{Ding20}); the first error terms consist of {\bf 1)} the Bayesian uncertainty of the relative reference position measured with respect to the primary in-beam calibrator, {\bf 2)} the position scatter of the primary in-beam calibrator (across all epochs) with respect to the off-beam phase calibrator, and {\bf 3)} the uncertainty of the off-beam phase calibrator position routinely updated in the Radio Fundamental Catalog\footnote{\url{astrogeo.org/rfc/}} (RFC).

\begin{table}
\caption{Inferred astrometric parameters and 68\% 
confidence intervals for \dnsa\ and \dnsb.}
\centering
\begin{tabular}{ccccccc}
\hline
\hline
reference\tablenotemark{a}  & $\mu_\alpha \equiv \dot{\alpha} \cos{\delta}$ & $\mu_\delta$ & $\varpi$ & $i_\mathrm{orb}$ & $\Omega_\mathrm{orb}$ & $\eta_\mathrm{EFAC}$ \\
 source  & (\maspy) & (\maspy) & (mas) & (deg) & (deg) &  \\
\hline
 \multicolumn{7}{c}{\dnsa} \\
\hline
\ibcaA   & 2.9(2) & $-5.6(4)$ & 0.30(10) & --- & --- & $1.1^{+0.7}_{-0.6}$ \\
\ibcaB   & 2.8(2) & $-6.0(4)$ & 0.25(12) & --- & --- & $2.7^{+1.4}_{-1.2}$ \\
VC & 2.8(2) & $-6.1(5)$ & 0.24(13) & --- & --- & $2.8^{+1.4}_{-1.0}$ \\
VC$+$\ibcaA & 2.9(1) & $-5.9(3)$ & 0.27(7) & --- & --- & $1.4^{+0.6}_{-0.5}$ \\
\hline
\multicolumn{7}{c}{\dnsb} \\
\hline
 \ibcbA   & 4.3(2) & $-5.2(4)$ & $0.29^{+0.12}_{-0.13}$ & $91^{+44}_{-43}$ & $237^{+117}_{-114}$ & $0.6^{+0.5}_{-0.4}$ \\
 \hline
\end{tabular}
\tablenotetext{a}{\raggedright The in-beam calibrators listed in Table~\ref{tab:sources} are referred to with their right ascensions. ``VC'' stands for the virtual calibrator described in Figure~\ref{fig:pointings} and Section~\ref{subsec:2Xtactic}.}
\tablenotetext{*}{\raggedright The absolute reference position of the DNSs are provided in Section~\ref{subsec:absolute_positions}.}
\label{tab:astrometric_results}
\end{table}

\section{Distances and Space Velocities}
\label{sec:D_and_Vt}

The astrometric determination of \dnsa\ and \dnsb\ considerably enriches the small sample of astrometrically studied field (i.e. not bound by a stellar cluster) DNSs.
So far, proper motion constraints have been reported for only 12 field DNSs (including the newly added \dnsa\ and \dnsb), which are summarized in Table~\ref{tab:D_and_Vt}. Among them, 6 DNSs also have parallax-based distance estimates (see the upper block of Table~\ref{tab:D_and_Vt}). In Table~\ref{tab:D_and_Vt}, we also provide useful information including the Galactic longitudes $l$, Galactic latitudes $b$, the spin periods $P_\mathrm{s}$, the orbital periods $P_\mathrm{b}$, the orbital eccentricities $e$ for the 12 DNSs, which were acquired from the {\tt PSRCAT} catalogue\footnote{\url{https://www.atnf.csiro.au/research/pulsar/psrcat/}} \citep{Manchester05}.
Besides, distances $d_\mathrm{DM}$ inferred from the dispersion measures (DMs) were derived with the {\tt PyGEDM} package\footnote{\label{footnote:pygedm}\url{https://github.com/FRBs/pygedm}} \citep{Price21} based on two models of Galactic free-electron distribution $n_\mathrm{e}$ --- the NE2001 model \citep{Cordes02} and the YMW16 model \citep{Yao17}. An indicative fractional uncertainty of 20\% is prescribed to each DM-based distance. 

\subsection{Double neutron star distances}
\label{subsec:DNS_distances}

Where parallax measurements are available, the parallax-based distances are adopted as the DNS distances $D$ (see Table~\ref{tab:D_and_Vt}). Otherwise, $D$ is approximated by the weighted mean of the two DM-based distances $d_\mathrm{DM}$, where the uncertainty of $D$ includes  the indicative 20\% fractional uncertainty as well as the standard deviation of the two $d_\mathrm{DM}$. We note that the DM-based distances might be subject to extra systematic uncertainties (see Section~\ref{sec:inidividual_systems}); nevertheless, they will get increasingly reliable with time, as $n_\mathrm{e}$ models become better constrained with growing number of independent pulsar distance measurements. 
The parallax-based distances of \dnsa\ and \dnsb\ were estimated in the same way detailed in Section~6 of \citet{Ding23}, 
while the parallax-based distances of the four other field DNSs with well constrained parallaxes are directly quoted from respective publications (see Table~\ref{tab:D_and_Vt} for the references).

\subsection{Transverse velocity estimation and caveats}
\label{subsec:v_t}

Investigating the kinematics of Galactic DNSs ideally requires the knowledge of the 3D space velocity, $v^\mathrm{birth}$, of each DNS system immediately after its birth (which is inherently linked to the formation history and the SN properties of the system, e.g. \citealp{tkf+17}). 
Here, ``space velocity'' refers to the peculiar velocity with respect to the local stellar neighbourhood of the DNS (see discussions in Section~6 of \citealp{Ding23}).
In theory, $v^\mathrm{birth}$ of a DNS system can be derived from its current full kinematic information (including the 3D velocity and the 3D position) and the age, $t_\mathrm{age}$, of the second-born NS, by applying a Galactic potential. This pathway to $v^\mathrm{birth}$ is, however, impractical, due to a highly uncertain value of $t_\mathrm{age}$ and the absence of the radial velocity, $v_\mathrm{r}$, for all known Galactic DNSs.  
With measured proper motion and 3D position (from sky coordinates and distance) of a DNS, we can unfortunately only constrain its current transverse velocity, i.e. the 2D velocity along the plane of the sky, and thus only infer its 2D peculiar velocity, $v_\perp$, relative to the stellar neighbourhood of the DNS system.

With no better option, this work uses $v_\perp$ as a proxy of $v_\perp^\mathrm{birth}$, i.e., the transverse velocity at birth, to probe the DNS kinematics. This approach hinges on the underlying assumption that $v_\perp$ is largely correlated with $v_\perp^\mathrm{birth}$, which is the first caveat to our analysis of DNS kinematics.
Conceivably, this assumption holds more truth for DNSs with low space velocities, as their space velocities change less over time in the Galactic potential.
Therefore, we expect that a $v_\perp$ distribution compiled from a large number of DNSs should be close to the DNS $v_\perp^\mathrm{birth}$ distribution at the low-space velocity end, while being more dispersed than the  $v_\perp^\mathrm{birth}$ distribution at the high-space velocity end.
Assuming natal kicks are equally possible in all directions (isotropic kicks), it is easy to calculate that on average $v_\perp^\mathrm{birth} = \left(\pi/4\right) \cdot v^\mathrm{birth}$. Accordingly, we approximate the $v^\mathrm{birth}$ distribution by $\left(4/\pi\right) \cdot v_\perp$ in this work.
To carefully justify the proposed $v_\perp$ to $v_\perp^\mathrm{birth}$ mapping (as well as the coefficient $4/\pi$) requires much more data and a dedicated population analysis, which is, however, not possible at this stage.

\begin{figure}[h]
    \centering
    \includegraphics[width=\textwidth]{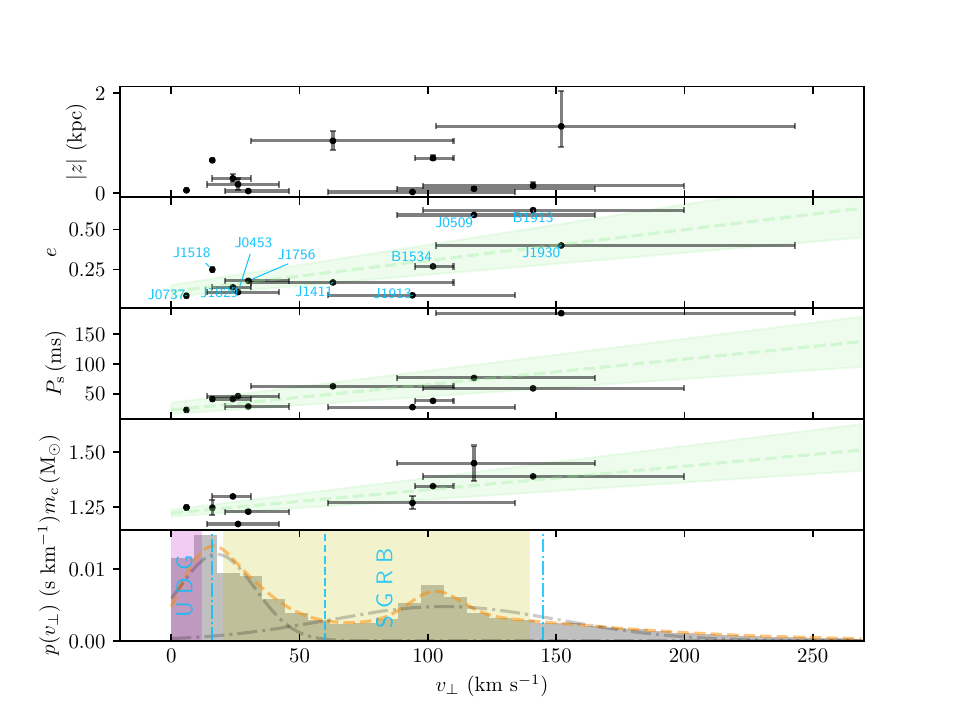}
    \caption{{\bf Top:} DNS transverse peculiar velocities, $v_\perp$, versus vertical distances, $|z|$, from the Galactic plane, from which no correlation is identified between $v_\perp$ and $|z|$. 
    {\bf Upper middle:} $v_\perp$ versus orbital eccentricity, $e$, where each DNS is marked with its right ascension. The best linear fit is given by: $e=1.9^{+0.6}_{-0.6}\times10^{-3}\cdot v_{\perp} + 0.11^{+0.04}_{-0.03}$ (green dashed line, while the green shade reflects the uncertainties).
    {\bf Central:} $v_\perp$ versus spin period, $P_\mathrm{s}$, of the first-born NS. The best linear fit is given by: $P_\mathrm{s}=0.43^{+0.11}_{-0.13}\;{\rm ms} \cdot v_\perp + 23^{+12}_{-7}\;{\rm ms}$.
    {\bf Lower middle:} $v_\perp$ versus companion mass $m_\mathrm{c}$. The best linear fit is given by: $m_\mathrm{c}=1.1^{+0.4}_{-0.3}\times10^{-3}\;{\rm M_\odot} \cdot v_\perp + 1.222^{+0.016}_{-0.016}\;{\rm M_\odot}$. In all the linear fitting relations of the 3 panels in the middle, $v_{\perp}$ is stated in ${\rm km\,s}^{-1}$.
    {\bf Bottom:} The probability density function (PDF) of $v_\perp$ approached by Monte Carlo re-sampling (see Section~\ref{subsec:v_t_distribution}). The vertical lines present the 16\%, 50\%, 84\% percentiles of the cumulative distribution function of $v_\perp$. Overlaid are the shaded $v_\perp$ ranges required for consistency with the observations of ultra-faint dwarf galaxies ($<\pi/4\cdot15$\,\kmps, see Section~\ref{subsubsec:implications_on_UFDGs}) and extragalactic short $\gamma$-ray bursts (20--140\,\kmps, see Section~\ref{subsubsec:consistency_with_SGRBs}).
    The dashed curve in dark orange shows the $v_\perp$ PDF smoothed with kernel density estimation (using {\tt scipy.stats.gaussian\_kde}, where the bandwidth is determined automatically), 
    while the two dash-dotted curves (in black) present the two Gaussian components fitted towards the smoothed $v_\perp$ PDF (see Section~\ref{subsec:v_t_distribution} for details). The two Gaussian components are plotted for the sole purpose of demonstration, as the unimodality of the current $v_\perp$ PDF cannot be ruled out owing to the small sample size of $v_\perp$ (see Section~\ref{subsubsec:bimodal_v_t}).}
    \label{fig:v_t}
\end{figure}

The $v_\perp$ of the DNSs were calculated in a consistent manner following Section~6 of \citet{Ding23}. The mathematical recipe of the calculation has been detailed in \citet{Verbunt17}.
We evaluated the $v_\perp$ uncertainties using Monte Carlo simulations, where proper motions and distances with asymmetric uncertainties were approached with split normal distributions \citep[e.g.][]{Possolo19}.
In the $v_\perp$ estimation, we approximated the motion of the stellar neighbourhood of each DNS by a circular Galactocentric motion parallel to the Galactic plane. This treatment is conceivably less reliable for DNSs $\gtrsim1$\,kpc away from the Galactic plane. Any anticorrelation (or correlation) between the Galactic height $|z|$ and $v_\perp$ would suggest that the approximation is oversimplified and inappropriate.
Though no correlation between $|z|$ and $v_\perp$ is identified (see the upper panel of Figure~\ref{fig:v_t}) with an insignificant Pearson correlation coefficient of $0.35^{+0.23}_{-0.24}$ (let alone a high false alarm probability of 0.29, see Table~\ref{tab:correlations}), we cautiously note that extra systematics of $v_\perp$ are likely introduced by the approximation for the relatively high-$|z|$ DNSs.
Due to the overly uncertain distance of PSR~J1757$-$1854 (see Table~\ref{tab:D_and_Vt}, or Figure~13 of \citealp{Cameron23}), we did not infer $v_\perp$ for PSR~J1757$-$1854. Neither did we use it in the sample studies discussed in Section~\ref{sec:discussion}. 
The other 11 DNSs used in the sample studies are hereafter referred to as the 11 DNSs.

\begin{table}
\caption{Distances and space velocities of astrometrically measured field double neutron stars}
\begin{minipage}{\textwidth}
\centering
\begin{tabular}{lccccccc}

\toprule
\toprule
PSR & $l$ & $b$ & $P_\mathrm{s}$ & $P_\mathrm{b}$ & $e$ & $m_\mathrm{p}$\,\tablenotemark{$f$} & $m_\mathrm{c}$\,\tablenotemark{$f$}  \\
& (deg) & (deg) & (ms) & (d) &  & (\Msun) & (\Msun) \\
 \midrule
 \dnsA\,\tablenotemark{$c_9$} & 168.3 & $-1.2$ & 76.5 & 0.38 & 0.59 & 1.36(8) & 1.45(8) \\
 J0737$-$3039A$\mid$B\,\tablenotemark{$c_1$} & 245.2 & $-4.5$ & 22.7$\mid$2773 & 0.10 & 0.088 & 1.33819(1) & 1.24887(1) \\

 J1518$+$4904\,\tablenotemark{$c_2$} & 80.8 & 54.3 & 40.9 & 8.6 & 0.25 & $1.470^{+0.030}_{-0.034}$ & $1.248^{+0.035}_{-0.029}$ \\
 B1534$+$12\,\tablenotemark{$c_{10}$} & 19.8 & 48.3 & 37.9 & 0.42 & 0.27 & 1.3330(2) & 1.3455(2) \\
 B1913$+$16\,\tablenotemark{$c_3$} & 50.0 & 2.1 & 59.0 & 0.32 & 0.62 & 1.438(1) & 1.390(1) \\
 \dnsB\,\tablenotemark{$c_{11}$} & 20.0 & $-16.9$ & 185.5 & 45.1 & 0.40 & $<1.32$ & $>1.30$  \\ 
\hline
J0453$+$1559\,\tablenotemark{$c_4$} & 184.1 & $-17.1$ & 45.8 & 4.1 & 0.11 & 1.559(5) & 1.174(4)  \\
J1411$+$2551\,\tablenotemark{$c_5$}  & 33.4 & 72.1 & 62.5 & 2.62 & 0.17 & $<1.62$ & $>0.92$ \\
J1756$-$2251\,\tablenotemark{$c_6$} & 6.5 & 0.95 & 28.5 & 0.32 & 0.18 & 1.341(7) & 1.230(7) \\

J1757$-$1854\,\tablenotemark{$c_{12}$} & 10.0 & 2.9 & 21.5 & 0.18 & 0.61 & 1.3412(4) & 1.3917(4) \\

J1829$+$2456\,\tablenotemark{$c_7$} & 53.3 & 15.6 & 41.0 & 1.18 & 0.14 & 1.306(4) & 1.299(4) \\
 J1913$+$1102\,\tablenotemark{$c_8$} & 45.3 & 0.19 & 27.3 & 0.21 & 0.090 & 1.62(3) & 1.27(3) \\
 \bottomrule
\end{tabular}\\

\vspace{1cm}

\begin{tabular}{lcccccc}

\toprule
\toprule
PSR & $\mu_\alpha$ & $\mu_\delta$   & $d_\mathrm{DM}^\mathrm{(NE)}$\,\tablenotemark{a} & $d_\mathrm{DM}^\mathrm{(YMW)}$\,\tablenotemark{a} & $D$\,\tablenotemark{b} & $v_\perp$  \\
& (\maspy) & (\maspy) & (kpc)  & (kpc) & (kpc) & (\kmps)  \\
 \midrule
 \dnsA\,\tablenotemark{$c_9$}  & 2.9(1) & $-5.9(3)$  & 1.9(4) & 1.6(3) & $4.2^{+1.6}_{-0.9}$ & $118^{+47}_{-30}$  \\
 J0737$-$3039A$\mid$B\,\tablenotemark{$c_1$} & $-2.57(3)$ & 2.08(4)  & 0.5(1) & 1.1(2) & 0.74(6) & 5.9(5)  \\
 J1518$+$4904\,\tablenotemark{$c_2$} & $-0.68(3)$ & $-8.53(4)$  & 0.6(1) & 1.0(2) & 0.81(2) & 16.0(6) \\
 B1534$+$12\,\tablenotemark{$c_{10}$} & 1.484(7) & $-25.29(1)$  & 0.9(2) & 0.9(2) & $0.94^{+0.07}_{-0.06}$ & $102^{+8}_{-7}$  \\
 B1913$+$16\,\tablenotemark{$c_3$} & $-0.77^{+0.16}_{-0.06}$ &  $0.01^{+0.10}_{-0.17}$  & 6(1) & 5(1) & $4.1^{+2.0}_{-0.7}$ & $141^{+59}_{-43}$  \\
 \dnsB & 4.3(2) & $-5.2(4)$  & 1.5(3) & 2.0(4) & $4.6^{+2.4}_{-1.4}$ & $152^{+91}_{-49}$  \\ 
\hline
J0453$+$1559\,\tablenotemark{$c_4$} & $-5.5(5)$ & $-6.0(4.2)$  & 1.1(2) & 0.5(1) & 0.6(4) & $26^{+16}_{-12}$  \\
J1411$+$2551\,\tablenotemark{$c_5$}  & $-3(12)$ & $-4(9)$  & 1.0(2) & 1.1(2) & 1.1(2) & $63^{+47}_{-32}$ \\
J1756$-$2251\,\tablenotemark{$c_6$} & $-2.42(8)$ & 0(2)  & 2.5(5) & 2.8(6) & 2.6(6) & $30^{+16}_{-9}$ \\

J1757$-$1854\,\tablenotemark{$c_{12},g$} & $-4.48(11)$ & $-0.5(12)$  & 7.4(1.5) & 19.5(3.9) & 9(7) & --- \\

J1829$+$2456\,\tablenotemark{$c_7$} & $-5.51(6)$ & $-7.8(1)$  & 1.2(2) & 0.9(2) & 1.1(3) & $24^{+7}_{-8}$  \\

J1913$+$1102\,\tablenotemark{$c_8$} & $-3.0(5)$ & $-8.7(1.0)$  & 7.6(1.5) & 7.1(1.4) & 7.3(1.5) & $94^{+40}_{-33}$  \\
 \bottomrule
\end{tabular}\\
\end{minipage}
\tablenotetext{a}{\raggedright $d_\mathrm{DM}^\mathrm{(NE)}$ and $d_\mathrm{DM}^\mathrm{(YMW)}$ represent the DM-based distances based on the NE2001 Galactic free electron distribution model \citep{Cordes02} and the YMW16 model \citep{Yao17}, respectively. All DM-based distances were acquired with the {\tt PyGEDM} package\textsuperscript{\ref{footnote:pygedm}}.
Indicative 20\% fractional uncertainties are applied to both $d_\mathrm{DM}^\mathrm{(NE)}$ and $d_\mathrm{DM}^\mathrm{(YMW)}$.}
\tablenotetext{b}{\raggedright For the upper block where the parallaxes have been constrained, parallax-based distances are reported and used for the calculation of the space velocities $v_\perp$. In particular, the distances of \dnsa\ and \dnsb\ were estimated in the same way as \citet{Ding23} (see Section~\ref{sec:D_and_Vt}). For any pulsar in the lower block where no parallax constraint is available yet, the distance is approximated by the weighted average of the two DM-based distances; the distance uncertainty consists of {\bf 1)} the weighted standard deviation of the two DM-based errors and {\bf 2)} the 20\% fractional uncertainty on the larger one of the two DM-based distances.}
\tablenotetext{c}{\raggedright References: 1. \citet{Kramer21a}, 2. \citet{Tan24,Ding23}, 3. \citet{Weisberg16,Deller18}, 4. \citet{Martinez15}, 5. \citet{Martinez17}, 6. \citet{Ferdman14}, 7. \citet{Haniewicz21}, 8. \citet{Ferdman20}, 9. \citet{Lynch18}, 10. \citet{Fonseca14,Ding23}, 11. \citet{Swiggum15}, 12. \citet{Cameron23}.}
\tablenotetext{f}{The $m_\mathrm{p}$ and $m_\mathrm{c}$ stand for, respectively, the first-born NS mass and the second-born NS mass. The mass uncertainties quoted for PSR~J1829$+$2456 components are half the 95.4\%-confidence-level uncertainties reported by \citet{Haniewicz21}.}
\tablenotetext{g}{PSR~J1757$-$1854 is not included in the sample studies of this paper due to the highly uncertain distance.}

\label{tab:D_and_Vt}
\end{table}

\section{Individual DNS systems}
\label{sec:inidividual_systems}

\subsection{\dnsa}
\label{subsec:J0509}

\dnsa\ was discovered in the Green Bank Northern Celestial Cap Pulsar Survey at 350\,MHz \citep{Stovall14}, and later identified as a DNS system in a highly eccentric ($e=0.586$) 9.1-hr orbit \citep{Lynch18}. The pulsar mass and the companion mass of the system are well determined thanks to the significant measurements of the periastron advance and the gravitational redshift \citep{Lynch18}. 
Among the 12 field DNSs having proper motion measurements, \dnsa\ possesses the second highest orbital eccentricity. Therefore, the relatively precise determination of $v_\perp$ for \dnsa\ (see Table~\ref{tab:astrometric_results}) is crucial for probing the $v_\perp$-to-$e$ correlation (see Section~\ref{subsec:corr}).
By comparing the parallax-based distance of \dnsa\ to the two DM distances, we find that, at $\approx2\,\sigma$ confidence, both DM distances (based on the NE2001 and the YMW16 $n_\mathrm{e}$ models) are underestimated.

\subsection{\dnsb}
\label{subsec:J1930}

\dnsb\ is the sixth pulsar discovered by students attending the Pulsar Search Collaboratory \citep{Rosen13} from the drift-scan observations made with the Green Bank Telescope at 350\,MHz \citep{Boyles13,Lynch13}. 
\dnsb\ is an exceptional DNS system: it challenges the DNS formation theory with the longest $P_\mathrm{s}$ (185.5\,ms) and the widest ($P_\mathrm{b}=45$\,d) orbit among the Galactic field DNSs (\citealp{Swiggum15}, also see Table~\ref{tab:D_and_Vt}).
Accordingly, the astrometric determination of \dnsb\ is important for investigating the $v_\perp$-to-$P_\mathrm{s}$ and $v_\perp$-to-$P_\mathrm{b}$ correlations (see Section~\ref{subsec:corr}).
Unfortunately, we did not achieve significant parallax measurement for \dnsb\ due to its radio weakness: random errors dominate the error budget of the position series. In this regard, the parallax precision of \dnsb\ can be potentially enhanced by a factor of $\gtrsim4$ in future with high-sensitivity VLBI arrays like the High Sensitivity Array.

\section{Sample studies of DNS kinematics}
\label{sec:discussion}

\subsection{Broad correlation analysis}
\label{subsec:corr}

The 3D systemic space velocity, $v^\mathrm{birth}$, at DNS birth, right after the second SN, is acquired from the combination of the anisotropic nature (i.e. natal momentum kick) and instantaneous mass loss of the SN producing the second-born NS. 
Therefore, it has been proposed that $v^\mathrm{birth}$ is likely correlated with a few parameters which include the orbital eccentricity, $e$, the orbital period, $P_\mathrm{b}$, and the companion mass $m_\mathrm{c}$ \citep[e.g.][]{tkf+17,Haniewicz21}.
Despite the caveats of $v_\perp$ estimation (see Section~\ref{subsec:v_t}), we investigate if $v_\perp$ is correlated with the aforementioned parameters. 
In addition to $v_\perp$, another important indicator of DNS kinematics is the DNS distance $|z|$ from the Galactic plane, which has also been refined with VLBI astrometry of DNSs. Assuming DNS systems were born in the Galactic plane and received natal kicks in random directions, DNSs with large space velocities are more likely found at high $|z|$.
To comprehensively examine the correlation between the two DNS kinematics indicators (i.e. $v_\perp$ and $|z|$) and other parameters, we carried out linear and power-law correlation analyses on 7 parameters, including $v_\perp$, $|z|$, $e$, $P_\mathrm{b}$, spin period (of the first-born and recycled NS) $P_\mathrm{s}$, $m_\mathrm{c}$ and pulsar mass, $m_\mathrm{p}$. The calculated Pearson correlation coefficients $\rho$ and their false alarm probabilities (or p-values) $P_\mathrm{F}$ are summarized in Table~\ref{tab:correlations}, where the two DNS systems (i.e., PSRs~J1930$-$1852 and J1411$+$2551) that have not had significant measurements of the pulsar mass are excluded from the calculation of mass-related Pearson correlation coefficients.

A correlation is normally considered tentative when $0.5\lesssim|\rho|\lesssim0.75$ and $P_\mathrm{F}<0.05$, while being strong if $\rho\gtrsim0.75$ and $P_\mathrm{F}<0.05$.
Unsurprisingly, two of the examined parameters (i.e., $P_\mathrm{s}$ and $|z|$) are well correlated with $P_\mathrm{b}$. These correlations are thought to have an origin in binary stellar evolution. For example, \citet{tlp15,tkf+17} demonstrated that the wider the pre-SN binaries orbit is, the less efficient the final Case~BB mass transfer to the first-born NS is (the reason being a shorter lifetime prior to the core collapse of more evolved helium star donors, i.e. the progenitors of the second-born NSs).
Therefore, after the second SN, wider-orbit DNS systems have, on average, slower spinning recycled NSs compared to more tight-orbit DNS systems. This is confirmed from our analysis which yields a Pearson correlation coefficient of $\rho=0.92$. Due to the unique kick magnitude and direction of each DNS system, some scatter is expected in the final correlation between $P_\mathrm{b}$ and $P_\mathrm{s}$.
\citet{tlp15,tkf+17} also argued for weak correlations between $e$ and both $P_\mathrm{b}$ and $P_\mathrm{s}$. 
We find that the inclusion of PSR~J1930$-$1852, which has exceptionally long $P_\mathrm{b}$ and $P_\mathrm{s}$ \citep{Swiggum15} among Galactic field DNSs, plays an essential role in dismissing the $e$-to-$P_\mathrm{b}$ correlation. But even without PSR~J1930$-$1852, a tentative $e$-to-$P_\mathrm{b}$ correlation is not revealed by our small sample. 
On the other hand, a tentative logarithmic correlation between $e$ and $P_\mathrm{s}$ is identified, even when PSR~J1930$-$1852 with the long $P_\mathrm{s}$ of 185.5\,ms is included.

Between the two kinematics indicators, $v_\perp$ and $|z|$, we did not find any significant correlation, which is the precondition of the $v_\perp$ estimation (see Section~\ref{subsec:v_t}). However, $v_\perp$ and $|z|$ apparently show correlations with the other parameters: $v_\perp$ is tentatively correlated with $e$ and $m_\mathrm{c}$ (see Figure~\ref{fig:v_t}), while $|z|$ is well correlated with $P_\mathrm{b}$. In addition, both kinematics indicators are correlated with $P_\mathrm{s}$ (see Figure~\ref{fig:v_t}). 
Large SN kicks will, in general, result in relatively large eccentricities. Hence, the former correlation between $v_\perp$ and $e$ is perhaps not surprising.
There are several components that affect the SN outcome: large kicks will disrupt the binaries. Because of the strong dependence on the stochastic kick direction in the second SN, as well as the different masses of the naked helium stars resulting from the previous common-envelope phase, it is non-trivial to directly link kinematic parameters ($v_\perp$ and $|z|$) of DNS systems to their orbital parameters and component masses.
A proposed correlation between the SN kick velocity magnitude and the resulting NS mass \citep{tkf+17} would translate into a (weaker) correlation between e.g. $v_\perp$ or $|z|$ and $m_\mathrm{c}$.
Our analysis confirms the postulated correlation between $v_\perp$ and $m_\mathrm{c}$, while finding no correlation between $|z|$ and $m_\mathrm{c}$.
On the other hand, any correlation between $|z|$ (or $v_\perp$) and $P_\mathrm{s}$ is somewhat unexpected and non-trivial to explain.

We note that great caution must be taken with postulating correlations involving parameters that evolve with time, such as $P_\mathrm{s}$ and $e$. To properly analyse the DNS population, time evolution must be included, e.g. \citet{tkf+17}. 
Future analysis with more DNS systems will likely improve the statistical significance of the tentative correlations suggested in this work.
Given no indication of insufficient $e$ or $P_\mathrm{s}$ coverage in the pulsar search programs (that led to the discoveries of the Galactic DNSs), no selection bias is uncovered for the $v_\perp$ distribution.
Among the four above-mentioned correlations between the two kinematics indicators and other parameters, linear correlations are either preferred over or comparable to the power-law counterparts by $\rho$ and $P_\mathrm{F}$ (see Table~\ref{tab:correlations}). This conclusion can also be drawn for the whole Table~\ref{tab:correlations}, with the \{$e, P_\mathrm{s}$\} pair being the only exception.

\begin{table}[htbp]
  \centering
  \caption{Pearson coefficients for {\bf (a)} linear correlation and {\bf (b)} logarithmic-scale correlation, and their false alarm probabilities}
  \vspace{0.3cm}
  \begin{minipage}[b]{\textwidth}
    \centering
    \begin{tabular}{c|cccccc}
\hline
\hline
& $e$ & $P_\mathrm{s}$ & $P_\mathrm{b}$ & $m_\mathrm{p}$ & $m_\mathrm{c}$ & $|z|$ \\
\hline
$v_\perp$ & \boldmath $\left.0.68^{+0.12}_{-0.15}\right\vert0.022$ & \boldmath $\left.0.65^{+0.17}_{-0.24}\right\vert0.029$ & $\left.0.45^{+0.23}_{-0.28}\right\vert0.16$ &  
$\left.0.12^{+0.18}_{-0.20}\right\vert0.76$ & \boldmath $\left.0.74^{+0.10}_{-0.15}\right\vert0.024$ & $\left.0.35^{+0.23}_{-0.24}\right\vert0.29$ \\
$|z|$ & $\left.0.12^{+0.05}_{-0.07}\right\vert0.72$ & \boldmath $\left.0.75^{+0.10}_{-0.18}\right\vert0.008$ & \boldmath $\left.0.78^{+0.11}_{-0.19}\right\vert0.005$ & $\left.-0.16^{+0.11}_{-0.11}\right\vert0.68$ & $\left.0.05^{+0.16}_{-0.12}\right\vert0.89$  & \\
$m_\mathrm{c}$ & \boldmath$\left.0.83^{+0.03}_{-0.06}\right\vert0.006$ & \boldmath$\left.0.72^{+0.07}_{-0.14}\right\vert0.03$ &  $\left.-0.40^{+0.12}_{-0.11}\right\vert0.29$ & $\left.-0.34^{+0.16}_{-0.14}\right\vert0.37$  & & \\
$m_\mathrm{p}$ & $\left.-0.19^{+0.14}_{-0.12}\right\vert0.62$ & $\left.-0.07^{+0.18}_{-0.17}\right\vert0.85$ &  $\left.0.32^{+0.09}_{-0.10}\right\vert0.40$  & & & \\
$P_\mathrm{b}$ & $0.20|0.55$ & \boldmath $0.92|0.00006$  & & & &  \\
$P_\mathrm{s}$ & $0.47|0.14$ &  & & & & \\
\end{tabular}\\
    \vspace{0.2cm}
    \textbf{(a)} 
    \label{tab:tableA}
  \end{minipage}\\

  \vspace{1cm}
  
  \begin{minipage}[]{\textwidth}
    \centering
        \begin{tabular}{c|cccccc}
\hline
\hline
& $\lg e$ & $\lg P_\mathrm{s}$ & $\lg P_\mathrm{b}$ & $\lg m_\mathrm{p}$ & $\lg m_\mathrm{c}$ & $\lg |z|$\\
\hline
$\lg v_\perp$ & \boldmath $\left.0.63^{+0.07}_{-0.08} \right\vert 0.040$ & \boldmath $\left.0.61^{+0.08}_{-0.10}\right\vert 0.047$ & $\left.0.15^{+0.09}_{-0.11}\right\vert 0.65$ & $\left.0.18^{+0.15}_{-0.17}\right\vert 0.64$ & $\left.0.63^{+0.10}_{-0.11}\right\vert 0.07$ & $\left.0.22^{+0.11}_{-0.12}\right\vert0.51$ \\
$\lg |z|$ & $\left.0.38(6)\right\vert0.25$ & \boldmath $\left.0.64^{+0.04}_{-0.06}\right\vert0.036$ & \boldmath $\left.0.75^{+0.04}_{-0.07}\right\vert0.008$ & $\left.-0.24^{+0.12}_{-0.13}\right\vert0.53$ & $\left.0.15^{+0.18}_{-0.15}\right\vert0.70$ &\\
$\lg m_\mathrm{c}$ & \boldmath$\left.0.78^{+0.05}_{-0.07}\right\vert0.013$ & $\left.0.63^{+0.06}_{-0.11}\right\vert0.07$ & $\left.-0.35^{+0.10}_{-0.09}\right\vert0.35$ &  $\left.-0.34^{+0.17}_{-0.15}\right\vert0.37$ & &\\
$\lg m_\mathrm{p}$ & $\left.-0.26^{+0.13}_{-0.12}\right\vert0.50$ & $\left.-0.02^{+0.17}_{-0.16}\right\vert0.95$ & $\left.0.28^{+0.07}_{-0.08}\right\vert0.47$ & & &\\
$\lg P_\mathrm{b}$ & $\left.0.21\right\vert0.54$ & \boldmath$\left.0.71\right\vert0.014$ & & & &\\
$\lg P_\mathrm{s}$ & \boldmath $\left.0.67\right\vert0.024$ &  & & & &\\
\end{tabular}\\
    \vspace{0.2cm}
    \textbf{(b)}
    \label{tab:tableB}
  \end{minipage}
  
\tablenotetext{*}{\raggedright The numbers on the left and right side of ``$|$'' are the Pearson correlations coefficients $\rho$ and the associated p-values $P_\mathrm{F}$, respectively. The uncertainties of the correlation coefficients reflect the errors of $m_\mathrm{p}$, $m_\mathrm{c}$ and $v_\perp$, and were estimated using Monte Carlo simulations. The p-values, or the false alarm probabilities of the correlation strengths, are calculated with the median $\rho$ of the simulations. $|\rho|>0.5$ with $P_\mathrm{F}<0.05$ are considered to suggest correlations, and are highlighted in bold.}
\label{tab:correlations}
\end{table}

\subsection{The transverse peculiar velocity distribution of field double neutron stars and its implications}
\label{subsec:v_t_distribution}

As noted in Section~\ref{subsec:v_t}, a $v_\perp$ distribution based on a large sample of astrometrically measured DNSs is expected to better approximate the $v_\perp^\mathrm{birth}$ distribution at the low velocity end, while being more dispersed than the $v_\perp^\mathrm{birth}$ counterpart at the high velocity end. With this in mind, we compiled the $v_\perp$ results in Table~\ref{tab:D_and_Vt} to a DNS $v_\perp$ distribution using Monte Carlo re-sampling. Specifically, we assume that all of the 11 field DNSs equally represent the whole population of field DNSs, and randomly drew 10,000 values of $v_\perp$ for each DNS from a split normal distribution. Subsequently, the $11\times10,000$ randomly drawn values of the DNSs were concatenated together. From this re-sampled $v_\perp$ chain, we 
estimated $v_\perp$ of field DNSs to be $60^{+85}_{-44}$\,\kmps (see Figure~\ref{fig:v_t}), hence the 3D space velocities of field DNSs should be $\sim4/\pi\times60$\,\kmps$=76$\,\kmps\ (see Section~\ref{subsec:v_t}). Here, the median of the $v_\perp$ chain is adopted as the $v_\perp$ estimate, while the central 68\% of the $v_\perp$ chain is used as the $v_\perp$ uncertainty interval. 
To mitigate binning effects, we smoothed the histogram of the $v_\perp$ simulations with kernel density estimation provided by {\tt scipy.stats.gaussian\_kde}. The resultant dashed curve (in the bottom panel of Figure~\ref{fig:v_t}) represents the estimated probability density distribution (PDF) of $v_\perp$.

\subsubsection{Consistency with extragalactic short $\gamma$-ray burst localizations}
\label{subsubsec:consistency_with_SGRBs}

As mentioned in Section~\ref{subsec:proper_motion_and_parallax_of_DNSs}, transverse velocities $v_\perp^\mathrm{SGRB}$ of SGRBs have been evaluated to be $\approx20$--140\,\kmps\ using projected SGRB displacements from the expected birth sites in their host galaxies, assuming that the SGRBs are associated with NS-NS mergers \citep{Fong13}.
It is timely to note that the $v_\perp^\mathrm{SGRB}$ estimation also has its caveats, hence being subject to extra systematic errors. For example, the estimation hinges on the reliability of the DNS delay time distribution. Besides, the distances to the SGRB host galaxies (and hence their projected sizes) are estimated with cosmological redshifts assuming negligible peculiar velocities.
Nevertheless, the $v_\perp^\mathrm{SGRB}$ inferred by \citet{Fong13} is highly consistent with our estimate of $v_\perp=60^{+85}_{-44}$\,\kmps\ (also see the bottom panel of Figure~\ref{fig:v_t}), which supports the idea that the observed extragalactic SGRBs are indeed driven by NS-NS mergers, and suggests that the underlying DNS delay time provided by \citet{Leibler10} is reasonable. The consistency between $v_\perp^\mathrm{SGRB}$ and $v_\perp$ strengthens the new research opportunity to be proposed in Section~\ref{subsec:prospects}.

\subsubsection{Implications on the r-process element production in ultra-faint dwarf galaxies}
\label{subsubsec:implications_on_UFDGs}

UFDGs are old and metal-poor satellite galaxies dominated by dark matter \citep{Simon07}. However, a small fraction of UFDGs are found to have exceptionally high abundance of r-process elements \citep{Ji16,Hansen17,Hansen20}, which is still poorly understood. If DNS mergers contributed to the r-process enrichment in UFDGs, the DNSs would either be gravitationally bound to the UFDG, or be able to merge within $\sim$1\,Myr (so that the produced r-process elements can potentially be recycled in the UFDG, \citealp{Safarzadeh19}).
As the latter possibility is disfavored by recent DNS delay time studies that show that most DNSs need $\gtrsim$10\,Myr to merge \citep{Andrews19a,Zevin22}, examining the former possibility with the DNS space velocity distribution becomes crucial for understanding the origin of the r-process enrichment in UFDGs.

To be gravitationally bound to a UFDG, DNSs are expected to have $v_\perp<\pi/4\cdot v_\mathrm{esc}^\mathrm{UFDG}$ on average, where the UFDG escape velocity $v_\mathrm{esc}^\mathrm{UFDG}$ is typically only 15\,\kmps\ \citep{Beniamini16}, and is limited to $\lesssim25$\,\kmps\ \citep{Safarzadeh19}.  
According to the current $v_\perp$ distribution, the probability that $v_\perp<\pi/4\cdot v_\mathrm{esc}^\mathrm{UFDG}$ is 11\% and 25\%, respectively, for $v_\mathrm{esc}^\mathrm{UFDG}=15$\,\kmps\ and $v_\mathrm{esc}^\mathrm{UFDG}=25$\,\kmps. In other words, $\sim$11\% of the DNSs born in UFDGs are expected to be gravitationally bound to the UFDGs. 
Here, we assume that the Galactic DNS $v_\perp$ distribution can well approximate the UFDG counterpart. This assumption might not be true due to the different properties of UFDGs and the Galaxy. 
In particular, the generally much lower metallicities in UFDGs \citep{Fu23} would reduce the minimum initial mass of progenitor stars required for CCSNe \citep{Han94}, and shift downward the mass range of progenitor stars that lead to ECSNe \citep{Podsiadlowski04}. Accordingly, it is likely that the DNS $v_\perp$ distribution in UFDGs is systematically lower than the Galactic counterpart. 
In this regard, $\sim$11\% might only be a lower limit to the fraction of DNSs gravitationally bound to the UFDGs.
The relation between the DNS space velocity distribution and the average metallicity of the host galaxy has not been well characterized, and is highly desired by our investigation (of the origin of the r-process elements in UFDGs using the kinematics of Galactic DNSs).
Due to the small sample size of $v_\perp$, the estimated fraction of DNSs bound to UFDGs is still indicative. 
Nevertheless, the estimation will become increasingly accurate when more astrometrically determined DNSs are available.

On the other hand, being bound to a UFDG does not guarantee r-process enrichment through DNS mergers: the DNSs are additionally required to merge within 1\,Gyr, so that the mergers precede the cessation of the star formation in UFDGs \citep{Brown14,Weisz14}. 
In addition, DNSs receiving the smallest kicks usually have smaller eccentricities, which tend to increase the merger timescales.
Therefore, we expect lower but still significant fraction of gravitationally bound DNSs eventually contributing to the r-process enrichment in UFDGs, given that the $\lesssim1$-Gyr merger time requirement is not hard to meet \citep{Zevin22}.

\subsubsection{A roadmap for testing multimodality of the $v_\perp$ distribution}
\label{subsubsec:bimodal_v_t}

It has been proposed that a $v_\perp$ distribution of a NS sub-group (e.g. millisecond pulsars or magnetars) can be used to probe the formation mechanism of that sub-group \citep[e.g.][]{tb96,Ding22,Ding23a,Ding23}.
The same scientific goal can be pursued with a DNS $v_\perp$ distribution.
Although the $v_\perp$ distribution we compile is based on only 11 isolated DNS systems, some global properties of the distribution may have emerged. In particular, an apparent bimodal feature can be seen in the estimated $v_\perp$ PDF (see the bottom panel of Figure~\ref{fig:v_t}).

To test the multimodality (including bimodality) of the sample of the 11 $v_\perp$ measurements, we employed Hartigan's dip test \citep{Hartigan85,Hartigan85a}, which starts from the null hypothesis that the $v_\perp$ distribution is unimodal, then examines the likelihood $P_\mathrm{Null}$ (also known as p-value) of the null hypothesis being true. 
In order to accommodate $v_\perp$ uncertainties, we carried out the dip test by the following procedure. We first re-sampled 10,000 sets of 11 $v_\perp$ (one $v_\perp$ per DNS per set), assuming split normal distribution for $v_\perp$ of each DNS. From every set, one p-value was acquired using the {\tt diptest}\footnote{\url{https://github.com/RUrlus/diptest}} python package (also see \url{https://cran.r-project.org/web/packages/diptest/index.html} for the earlier {\tt R} package).
Finally, the mean of the 10,000 p-values is adopted as the $P_\mathrm{Null}$ estimate.
In this way, we obtained $P_\mathrm{Null}=50$\%, which suggests that no conclusion can be drawn with regard to the  multimodality of the current $v_\perp$ distribution.

The large $P_\mathrm{Null}$ is not surprising given the small sample size (of 11).
If the observed $v_\perp$ PDF (in Figure~\ref{fig:v_t}) as well as the bimodality is true, we find that about 120 and 160 extra astrometrically determined DNSs are required to rule out unimodality of the $v_\perp$ sample at 90\% and 95\% confidence, respectively. This prospect (of probing multimodality of the $v_\perp$ distribution) might be realized in the next decade with the ongoing and future pulsar search programs utilizing high-sensitivity pulsar timing facilities such as FAST \citep[e.g.][]{Li16,Han21,Miao23}, MeerKAT \citep[e.g.][]{Bailes20,Kramer21}, ngVLA \citep[e.g.][]{Murphy18,McKinnon19} and the SKA \citep[e.g.][]{Dewdney09,Keane+15}. 
If unimodality of the $v_\perp$ distribution can be ruled out at high confidence, one can proceed to decompose the $v_\perp$ PDF into multiple modes, and investigate the likely formation channel(s) associated with each mode. 
For example, if in future $\sim$160 isolated DNSs were found to follow a bimodal $v_\perp$ distribution similar to the one in Figure~\ref{fig:v_t}, a bimodal Gaussian distribution could be fitted against the future $v_\perp$ PDF. For the only purpose of demonstration, the best-fit Gaussian components for the current $v_\perp$ PDF (estimated in this work) are illustrated with the two dash-dotted curves in the bottom panel of Figure~\ref{fig:v_t}.

\section{Conclusions and future prospects}
\label{sec:conclusions_and_prospects}

\subsection{Conclusions}
\label{subsec:conclusions}

In this work, we measured proper motions and parallaxes for two DNS systems PSRs~\dnsA\ and \dnsB\ with VLBA observations. In a consistent manner, we estimated the two kinematic indicators, the transverse space velocity $v_\perp$ and the vertical height $|z|$, for all Galactic DNSs astrometrically measured (see Section~\ref{sec:D_and_Vt}). With this small sample of 11, we investigated the correlation between the kinematic indicators and 5 other parameters, including the orbital period $P_\mathrm{b}$, the spin period $P_\mathrm{s}$, the orbital eccentricity $e$, the companion mass $m_\mathrm{c}$ and the pulsar mass $m_\mathrm{p}$. Tentative correlations were identified between $v_\perp$ and $e$, $v_\perp$ and $P_\mathrm{s}$, $v_\perp$ and $m_\mathrm{c}$, $|z|$ and $P_\mathrm{s}$, and $|z|$ and $P_\mathrm{b}$ (see Section~\ref{subsec:corr}).
By combining the $v_\perp$ estimates, we obtained the preliminary $v_\perp$ distribution of DNSs (see Section~\ref{subsec:v_t_distribution}), which is used to pursue three scientific studies. Firstly, we found that the $v_\perp$ distribution is consistent with the indicative transverse velocities $v_\perp^\mathrm{SGRB}$ of short gamma-ray bursts (SGRBs) evaluated by \citet{Fong13}, which suggests that the underlying DNS delay time provided by \citet{Leibler10} is reasonable. Secondly, according to the $v_\perp$ distribution and the typical escape velocity of ultra-faint dwarf galaxies (UFDGs), $\gtrsim11$\% of the DNSs born in UFDGs are gravitationally bound to UFDGs. We argue that a significant fraction of the DNSs gravitationally bound to UFDGs would eventually contribute to the r-process enrichment of the UFDGs. Therefore, DNS mergers cannot be ruled out as the source of r-process enrichment in UFDGs. Finally, we looked into the prospects of using the $v_\perp$ distribution to probe the formation channels of DNSs. Our simulation suggests that astrometric measurements of $\sim120$ additional isolated DNSs were needed to rule out unimodality of the $v_\perp$ distribution at 90\% confidence, if the $v_\perp$ distribution shown in Figure~\ref{fig:v_t} is true.

\subsection{Future prospects}
\label{subsec:prospects}

\begin{enumerate}[label=(\roman*)]
    
    \item The availability of the three components --- the $v_\perp$ distribution (directly approached with Galactic DNSs), the SGRB displacements from the expected birth sites in the host galaxies \citep[e.g.][]{OConnor22,Fong22}, the DNS delay time distribution (constrained by Galactic DNSs, SGRB host galaxies and GW events, e.g. \citealp{Andrews19a,Beniamini19a,Zevin22}), promises a new research opportunity: by combining the knowledge of the three components, one can uncover the selection biases (in the three components) to unprecedented details, and achieve the most reliable posterior distributions for the three components. 
    
    \item As mentioned in Section~\ref{subsubsec:bimodal_v_t}, ongoing and future relativistic binary pulsar search programs using high-sensitivity radio telescopes (or tied arrays) will drastically increase the number of Galactic DNSs, which will eventually enable the multimodality test of the $v_\perp$ distribution, and refine the estimation of the fraction of DNSs gravitationally bound to UFDGs (see Section~\ref{subsubsec:implications_on_UFDGs}). 
    Meanwhile, high-sensitivity VLBI observations of DNSs will not only increase the number of DNSs well measured astrometrically, but also achieve higher astrometric precision for the DNSs.
    
\end{enumerate}

\section*{Acknowledgments}
HD acknowledges the EACOA Fellowship awarded by the East Asia Core Observatories Association.
JKS, RSL, and SC are members of the NANOGrav Physics Frontiers Center, which is supported by NSF award PHY-1430284.
This work is based on observations with the Very Long Baseline Array (VLBA), which is operated by the National Radio Astronomy Observatory (NRAO). The NRAO is a facility of the National Science Foundation operated under cooperative agreement by Associated Universities, Inc.
This work made use of the Swinburne University of Technology software correlator, developed as part of the Australian Major National Research Facilities Programme and operated under license \citep{Deller11a}.

\software{{\tt numpy} \citep{Harris20}, 
{\tt scipy} \citep{Virtanen20}, 
{\tt astropy} \citep{Astropy-Collaboration13,Astropy-Collaboration18,Astropy-Collaboration22}, {\tt matplotlib} \citep{Hunter07}, {\tt bilby} \citep{Ashton19}, {\tt psrqpy} \citep{Pitkin18}, {\tt adjustText}\footnote{\url{https://github.com/Phlya/adjustText}} \citep{Flyamer24}}

\bibliographystyle{aasjournal}
\bibliography{refs,more-refs}

\end{document}